\documentclass[preprint,showkeys,showpacs,10pt,prd]{revtex4}%

\makeatletter

\renewcommand*{\p@subsection}{}

\renewcommand*{\p@subsubsection}{}
\makeatother

\usepackage[Export]{adjustbox}
\usepackage{amsmath, amssymb, amsfonts, amsthm}
\usepackage{epsfig}
\usepackage{graphicx}  
\usepackage[utf8]{inputenc}
\usepackage{color}     
\usepackage{lscape}
\usepackage{color}

\begin{document}

\title{Matter-antimatter asymmetry induced by non-linear electrodynamics}

\author{H. B. Benaoum
} 
\email{hbenaoum@sharjah.ac.ae}
\affiliation{
Department of Applied Physics and Astronomy, \\
University of Sharjah, United Arab Emirates 
}
\author{A. {\"O}vg{\"u}n}
\email{ali.ovgun@emu.edu.tr}
\homepage[]{https://aovgun.weebly.com/}

\affiliation{Physics Department, Eastern Mediterranean University, Famagusta, 99628 North Cyprus via Mersin 10, Turkey.}

\date{\today}
\begin{abstract}
In this work, we propose an economical model to address some open cosmological problems such as the absence of the initial cosmological singularity, an early acceleration of the Universe and the generation of matter-antimatter asymmetry. The model is based on a scenario in which the early Universe consists of a non-linear electrodynamics fields. It is found that the non-linear electrodynamics model has an equation of state $p=\frac{1}{3} \rho - \frac{4}{3} \beta \rho^{1+\alpha}$ which shows that the Universe undergoes an early epoch acceleration to a radiation era given by $p =\frac{1}{3} \rho$. We show that the singularities in the energy density, pressure and curvature are absent at early stages. In our scenario, the baryon asymmetry is generated by the non-linearity parameter $\beta$. Additionally, we calculate the resulting baryon asymmetry and discuss how a successful gravitational baryogenesis is obtained for different values of the model's parameter space.
\end{abstract}


\maketitle

\section{Introduction} 
The baryon asymmetry, which is an excess of matter over antimatter in the visible Universe, is one of the unsolved problem in cosmology. It is typically characterized by the dimensionless quantity: 
\begin{eqnarray}
\eta_B & = & \frac{n_B - n_{\bar{B}}}{s_{\gamma}}
\end{eqnarray}
where $n_{B (\bar{B})}$ is the number of baryons (anti-baryons) density and $s_{\gamma}$ is the radiation entropy density. According to observations of the Cosmic Microwave Background radiation (CMB) \cite{ade2015}, predictions of Big-Bang Nucleosynthesis (BBN) \cite{aghanim2018} and the absence of intense radiation from matter-antimatter annihilation \cite{cohen1998}, the observed baryon number asymmetry today is:
\begin{eqnarray}
\eta_B & = &  \left(6.04 \pm 0.08 \right) \times 10^{-10}
\end{eqnarray}
To explain the observed baryon asymmetry, Sakharov \cite{sakharov1967} has postulated that any local CPT invariant quantum field theory (QFT) must have interactions which violate the conservation of baryon number, the charge and the charge-parity (CP) symmetries. There should also be a departure from thermal equilibrium, since in a CPT invariant theory in thermal equilibrium any Baryon number violation will be washed out.  
Various baryogenesis models, based on electroweak baryogenesis have been proposed which satisfy the well known conditions for baryogenesis advanced by Sakharov. In this paper, we consider a scenario termed 
Gravitational Baryogenesis (GB) \cite{davoudiasl2004}, which can occur naturally in an effective theory of gravity. It introduces an effective coupling between the derivative of the Ricci scalar curvature $R$ and the baryon current $J_B^{\mu}$. Such a term attributes opposite sign to the energy contributions of particles and anti-particles, thus violating CPT symmetry in an expanding Universe. \\
For an Einstein-Hilbert gravity in a Friedmann-Robertson-Walker (FRW) background, the Ricci scalar curvature $R$ can be easily deduced from the Einstein equations as: 
\begin{eqnarray}
R & = & 6 \left( \frac{\ddot{a}}{a}+ \left(\frac{\dot{a}}{a} \right)^2 \right) = \kappa^2 \left( \rho - 3 p \right)
\end{eqnarray}
where $a$ is the scale factor, $\rho$ and $p$ are the energy density and pressure of the cosmic fluid, $\kappa^2 = 8 \pi G =M_p^{-2}$ and $M_p\simeq 2.4 \times 10^{18} ~GeV$ is the reduced Planck mass. It is easy to notice that the observed baryon asymmetry cannot be generated from GB when $p = \frac{1}{3} \rho$ which corresponds to a radiation dominated Universe. The GB mechanism has been studied in the context of  $f (R)$
gravity \cite{lambiase2006,ramos2017}, brane world scenarios \cite{shiromizu2004}, loop quantum cosmology \cite{odintsov2016}, $f (T )$ gravity \cite{oikonomou2016}, Gauss–Bonnet brane world cosmology \cite{bento2005}, Gauss–Bonnet gravity \cite{odintsov20162}, running vacuum models \cite{lima2017,pan2017}, Symmergent Gravity \cite{Cimdiker:2020enx,Demir:2019imw,Demir:2016ubi,Azri:2018qux,Azri:2017uor}, alternative models \cite{Hernandez-Almada:2018osh,Vazquez:2018qdg,Sert:2019qfl,Dil:2020fvf,Oztas:2018jsu} and Horava gravity \cite{maity2019}. \\
In their seminal paper in 1934 \cite{born1934}, Born and Infeld have proposed the non-linear electrodynamics (NLED) to cure the divergence of self-energy of charged particles. Other non-linear generalizations of classical electrodynamics have been introduced in the literature by many authors \cite{gitman2014}-\cite{Otalora:2018bso}. Such theories, in particular the Born-Infled theory turns out to play an important role in string theory. In recent years NLED has been revived and has attracted a significant amount of interest in cosmological models \cite{vollick2003}-\cite{zorin2002}, black holes, wormholes, gravitational baryogenesis \cite{mosquera2009} and astrophysics. 
The standard cosmological model involving Maxwell electrodynamics based on the FRW geometry leads to a space-like singularity in the past.  It is widely believed that the Universe in the early era, which was dominated by radiation, had a very strong and highly non-linear electromagnetic field. Such a non-linear electromagnetic field coupled to gravitational field can explain the acceleration (i.e inflation) of the Universe at early stages. In addition, in some models of NLED the initial singularity is absent. \\

In this paper, we present an economical model that could account for the absence of the initial cosmological singularity, inflation and matter antimatter asymmetry. In section 2, we introduce our phenomenological model which is based on a new Lagrangian for NLED depending on two real parameters $\alpha$ and $\beta$. We demonstrate in section 3 and 4 that the Universe has no Big-Bang singularity and tends to accelerate at early era. Section 5 is devoted to GB in the framework of NLED. We investigate the qualitative implications of NLED by calculating the corresponding baryon asymmetry and compare our results with the observational data. Some conclusions are presented in section 6. \\
\section{Non-linear Electrodynamics}
In non-linear electrodynamics, the classical Maxwell Lagrangian density is replaced by:
\begin{eqnarray}
{\cal L}_{nled} & \equiv & -{\cal F}  ~f \left( {\cal F} \right), 
\label{lag}
\end{eqnarray}
where ${\cal F} = \frac{1}{4} F_{\mu \nu} F^{\mu \nu} = \frac{1}{2} (B^2 - E^2)$ and $f \equiv f ({\cal F})$ is a functional depending on the field strength. 
The energy-momentum tensor for a Lagrangian density ${\cal L}_{nled}$, is: 
\begin{eqnarray}
T^{\mu \nu} & = & H^{\mu \lambda} F^{\nu}_{\lambda} - g^{\mu \nu} {\cal L}_{nled} 
\end{eqnarray}
where 
\begin{eqnarray}
H^{\mu \lambda} & = & \frac{\partial {\cal L}_{nled}}{\partial F_{\mu \lambda}} = \frac{\partial {\cal L}_{nled}}{\partial {\cal F}} F^{\mu \lambda} 
\end{eqnarray}
For a Lagrangian density given by (\ref{lag}), the energy-momentum is:
\begin{eqnarray}
T^{\mu \nu} & = & - \left( f + {\cal F} \frac{d f}{{d \cal F}} \right)  F^{\mu \lambda} F^{\nu}_{\lambda} + 
g^{\mu \nu} {\cal F} f 
\end{eqnarray}
where the energy density $\rho$ and pressure $p$ can be found to be:
\begin{eqnarray}
\rho & = & {\cal F} f - E^2 \left( f + {\cal F} \frac{d f}{d {\cal F}}\right) \nonumber \\
p & = & - {\cal F} f + \frac{2 B^2 - E^2}{3} \left(f + {\cal F} \frac{d f}{d {\cal F}}  \right)
\end{eqnarray}
One can assume that the curvature is much larger than the
wavelength of the electromagnetic waves, because the electromagnetic fields
are the stochastic background. The average of the electromagnetic fields that are sources
in general relativity have been used to obtain the isotropic FRW space-time \cite{tolman}.
For this reason, one uses the average values of the electromagnetic fields as follows 
\begin{equation}
\langle \mathbf{E}\rangle =\langle \mathbf{B}\rangle =0,\text{ }\langle
E_{i}B_{j}\rangle =0,
\end{equation}%
\begin{equation*}
\langle E_{i}E_{j}\rangle =\frac{1}{3}E^{2}g_{ij},\text{ }\langle
B_{i}B_{j}\rangle =\frac{1}{3}B^{2}g_{ij}.
\end{equation*}%
Note that later we omit the averaging brackets $\langle $ $\rangle $
for simplicity. The most interesting case of this method occurs only when the average of the magnetic field is not zero. \cite{tolman}. The universe has a magnetic property that the magnetic field is frozen in the cosmology where the charged primordial plasma screens the electric field. 
The non-linear electrodymanics is expected to play a crucial role in the evolution of the Universe \cite{gitman2014}-\cite{mosquera2009}. For this purpose, we propose the following non-linear electrodynamics Lagrangian in which the functional $f$ is purely magnetic (i.e. $\vec{E} = \vec{0}$) depending on two real parameters $\alpha$ and $\beta$,
\begin{eqnarray}
f \left({\cal F} \right) & = & \frac{{1}}{\left( \beta {\cal F}^{\alpha} + 1 \right)^{1/\alpha}},
\end{eqnarray}
where $\beta {\cal F}^{\alpha}$ is dimensionless and for $\beta =0$, ${\cal L} (f)= 1$ which is the usual electrodynamics Lagrangian. We will see later on that the non-linear parameter $\beta$ is related to the initial energy of the magnetic field and $\alpha$ has to be positive in order to have an early acceleration of the Universe  (negative values for $\alpha$ correspond to a late acceleration of Universe which will not be considered here). The energy density and pressure becomes:
\begin{eqnarray} 
\rho & = & \frac{{\cal F}}{\left( \beta {\cal F}^{\alpha} + 1 \right)^{1/\alpha}} \nonumber \\
p & = & - \frac{{\cal F}}{\left( \beta {\cal F}^{\alpha} + 1 \right)^{1/\alpha}} + \frac{2}{3} \frac{B^2}{\left( \beta {\cal F}^{\alpha} + 1 \right)^{1+ 1/\alpha}} 
\end{eqnarray}
where ${\cal F} = \frac{1}{2} B^2$. Also, one can write

\begin{equation}
 \rho+p=\frac{4}{3} \frac{\cal F}{\left( \beta {\cal F}^{\alpha} + 1 \right)^{1 + 1/\alpha}}, \label{rp2}
\end{equation}
and
\begin{equation}
 \rho+3p= \frac{2 {\cal F} \left(1 - \beta {\cal F}^{\alpha} \right)}{\left( \beta {\cal F}^{\alpha} + 1 \right)^{1 + 1/\alpha}}  . \label{rp3}
\end{equation}
The equation of state satisfied by the above nonlinear electrodynamics Lagrangian density is: 
\begin{eqnarray}
p & = & \frac{1}{3} \rho \left(1 - 4 \beta \rho^{\alpha} \right)  ~~~. 
\label{eos1}
\end{eqnarray}
We notice that for $\beta=0$, the above equation of state reduces to the Maxwell radiation EoS. \\
\section{Non-linear electrodynamics Cosmology}
The four-dimensional Einstein-Hilbert action of gravity coupled to a non-linear electromagnetic field is given by:
\begin{eqnarray}
S & = & \int d^4 x \sqrt{-g} \left( \frac{1}{2 \kappa^2} R + {\cal L}_{nled} \right)
\end{eqnarray}
where $R$ is the Ricci scalar and $\kappa^{-1}= M_p$ is the reduced Planck mass. The Einstein fields equation is derived from the action 
\begin{equation}
R_{\mu \nu }-\frac{1}{2}g_{\mu \nu }R=-\kappa ^{2}T_{\mu \nu },  \label{EQ}
\end{equation}
and $\ $%
\begin{equation}
\partial _{\mu }\left( \sqrt{-g}\frac{\partial \mathcal{L}_{nled}}{\partial 
\mathcal{F}}F^{\mu \nu }\right) =0.
\end{equation}

A homogeneous and isotropic flat Universe is described by the FRW metric: 

\begin{equation}
ds^{2}=-dt^{2}+a(t)^{2}\left[ dr^{2}+r^{2}\left( d\theta ^{2}+sin(\theta
)^{2}d\phi ^{2}\right) \right] 
\end{equation}%
where $a$ is the expansion scale factor. With the help of the FRW metric, one obtains the Friedmann equations as follows: 
\begin{eqnarray}
H^2 & = & \left( \frac{\dot{a}}{a} \right)^2 = \frac{\kappa^2}{3} ~\rho \nonumber \\
3 \frac{\ddot a}{a} & = & - \frac{\kappa^2}{2} \left( \rho + 3 p \right)
\end{eqnarray}
where $H$ is the Hubble parameter.  \\
By using the conservation of the energy-momentum tensor $\nabla ^{\mu }T_{\mu \nu }=0$, the continuity equation  is derived as $\dot{\rho} + 3 H (\rho + p) =0$, then one can find a relation between the electromagnetic field ${\cal F}$ and the scale factor $a$ as: 
\begin{eqnarray}
{\cal F} & = & {\cal F}_I \left(\frac{a_I}{a} \right)^4 = 2 B^2 
\end{eqnarray}
It follows that:
\begin{eqnarray}
B & = & B_I ~\left(\frac{a_I}{a} \right)^2
\end{eqnarray}
where $a_I$ is the scale factor at inflation and ${\cal F}_I = 2 B_I{^2}$. \\
From the above equation, we find the energy density and the pressure in terms of the scale factor $a$,
\begin{eqnarray}
\rho & = & \frac{\rho_I}{(\gamma + (1- \gamma) (\frac{a}{a_I} )^{4 \alpha} )^{1/\alpha}} \nonumber \\
p & = & \rho_I \frac{ -\gamma + \frac{1}{3} (1- \gamma)(\frac{a}{a_I})^{4 \alpha}}{
(\gamma + (1- \gamma) (\frac{a}{a_I} )^{4 \alpha} )^{1+1/\alpha}} \label{eos}
\end{eqnarray}
where $\rho_I = \rho (a_I)$ is the energy density at scale $a_I$ and $\gamma = \beta \rho_I^{\alpha}$ is a dimensionless parameter encoding the non-linearity. \\

The energy density $\rho $ and the pressure $p$ in the limit of $a(t)\rightarrow 0$  and $a(t)\rightarrow \infty $ become:
\begin{equation}
\lim_{a(t)\rightarrow 0}\rho (t)=\beta^{-1/\alpha},~~\lim_{a(t)\rightarrow
0} p(t)=-\beta^{-1/\alpha},  \label{15}
\end{equation}%
\begin{equation}
~~\lim_{a(t)\rightarrow \infty }\rho (t)=\lim_{a(t)\rightarrow \infty
}p(t)=0.  \label{166}
\end{equation}%
It is interesting to note that the non-linearity $\beta$ is related to the energy density at the early stage of the Universe 
\begin{eqnarray} 
\rho_0 & = & \rho ( a=0) = \beta^{-1/\alpha}
\end{eqnarray}
and the dimensionless parameter $\gamma$ becomes,
\begin{eqnarray} 
\gamma & = & \left( \frac{\rho_I}{\rho_0}\right)^{\alpha}
\end{eqnarray}

It follows from (\ref{15}) that the singularity of the energy density and the pressure is absent due to the non-linearity. The Ricci scalar, which represents the curvature of space-time, is calculated by using Einstein's field equation  and the energy-momentum tensor, 
\begin{equation}
R=\kappa ^{2}(\rho -3p).  \label{16r}
\end{equation}%
The Ricci tensor squared $R_{\mu \nu }R^{\mu \nu }$and the Kretschmann
scalar $R_{\mu \nu \alpha \beta }R^{\mu \nu \alpha \beta }$ are also
obtained as 
\begin{equation}
R_{\mu \nu }R^{\mu \nu }=\kappa ^{4}\left( \rho ^{2}+3p^{2}\right) ,
\label{rrrr}
\end{equation}%
\begin{equation}
R_{\mu \nu \alpha \beta }R^{\mu \nu \alpha \beta }=\kappa ^{4}\left( \frac{5%
}{3}\rho ^{2}+2\rho p+3p^{2}\right) .
\end{equation}%
Similarly, it is easy to see that the Ricci scalar curvature, the Ricci tensor and the Kretschmann
scalar are not singular, 
\begin{equation}
\lim_{a(t)\rightarrow 0}R(t)=4 \kappa^2 \rho_0,
\end{equation}%
\begin{equation}
\lim_{a(t)\rightarrow 0}R_{\mu \nu }R^{\mu \nu }=4 \kappa^2 \rho_0^2%
,
\end{equation}%
\begin{equation}
\lim_{a(t)\rightarrow 0}R_{\mu \nu \alpha \beta }R^{\mu \nu \alpha \beta }=\frac{8 \kappa^4 \rho_0^2}{3}%
,  \label{18}
\end{equation}%
\begin{equation}
\lim_{a(t)\rightarrow \infty }R(t)=\lim_{a(t)\rightarrow \infty }R_{\mu \nu
}R^{\mu \nu }=\lim_{a(t)\rightarrow \infty }R_{\mu \nu \alpha \beta }R^{\mu
\nu \alpha \beta }=0.
\end{equation}%
The absence of singularities is an attractive feature which is peculiar of NLED. 

The second Friedmann equation: 
\begin{equation}
3 \frac{\ddot a}{a} = - \frac{\kappa^2}{2} \left( \rho + 3 p \right) \label{fr2}
\end{equation} describes the universe's rate of expansion, how quickly the expansion is speeding up or slowing down. The most important condition for the accelerated universe is $\rho+3p<0$. 
Using the equation (\ref{eos}), we find that equation of state parameter is: 
\begin{eqnarray}
\omega & = & \frac{p}{\rho} = \frac{ -1 + \frac{1}{3 \gamma} (1- \gamma) (\frac{a}{a_I})^{4 \alpha}}{1 + \frac{1}{\gamma} (1- \gamma) (\frac{a}{a_I})^{4 \alpha} }
\end{eqnarray}
It follows that for $\alpha>0$, 
\begin{itemize}
\item{at small scale $ a <<a_I$} \\
\begin{eqnarray}
\omega & = & \frac{p}{\rho} \simeq -1 + \frac{4}{3} (\frac{1}{\gamma} - 1) ~\left(\frac{a}{a_I} \right)^{4 \alpha} 
\end{eqnarray}
\item{at large scale $ a >>a_I$} \\
\begin{eqnarray}
\omega & = & \frac{p}{\rho} \simeq \frac{1}{3} - \frac{4}{3} \left(\frac{\gamma}{1- \gamma} \right) ~\left(\frac{a_I}{a} \right)^{4 \alpha} ~~~~~~`.
\end{eqnarray}
\end{itemize}
It shows that the Universe has a negative equation of state for small $a$ and radiation dominated Universe for large $a$. 
\section{Evolution of the Universe}
The first Friedmann equation gives,
\begin{eqnarray}
H^2 & = & \left( \frac{\dot{a}}{a} \right)^2 = \frac{\kappa^2}{3} ~\rho \label{fr1}.
\end{eqnarray}
A general solution of the above equation in terms of the Hubble parameter is obtained as:
\begin{widetext}

\begin{eqnarray}
t + const & = &  
\frac{1}{2 H} ~{_2}F_1 \left(1,-\frac{1}{2 \alpha}; 1 - \frac{1}{2 \alpha}; \gamma  \left( \frac{H}{H_I} \right)^{2 \alpha}\right)
\end{eqnarray}
\end{widetext}
where $H_I = H (a_I)$ is the Hubble constant at inflation. In the Figures 1 and 2, the parameter $H_I (t + const)$ is plotted as a function of the scale factor $a/a_I$ and the Hubble parameter $H/H_I$ for different values of the $\gamma$ and $\alpha$. It is worth noticing that $\gamma=0$ corresponds to a radiation dominated Universe where $t \propto a^2 \propto 1/2H$ as shown in figures 1 and 2. For $\gamma \neq 0$, we see that the Universe has no initial singularity.


\begin{figure}[htp]
\centering
  \includegraphics[width=.6\textwidth]{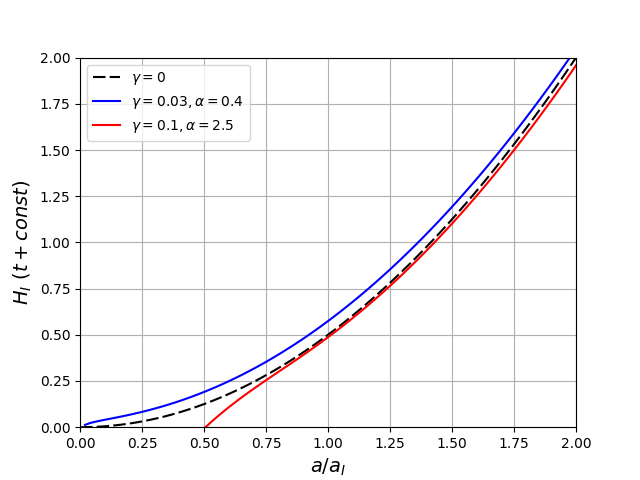} \caption{$H_I (t+ const)$ versus scale factor $a/a_I$ for different values of $\gamma$ and $\alpha$} \end{figure}

\begin{figure}[htp]
\centering
  \includegraphics[width=.6\textwidth]{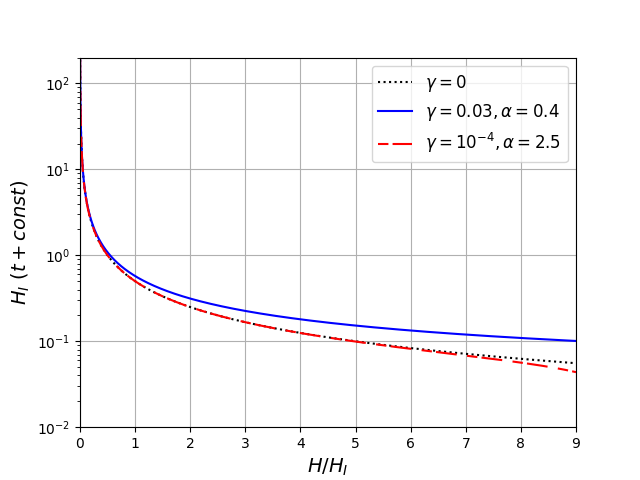} 
\caption{
$H_I (t + const)$ versus $H/H_I$ for different values of $\gamma$ and $\alpha$.Note the logarithmic scale for the vertical axis}
\end{figure}

Now by using both Friedmann equations (\ref{fr1}) and (\ref{fr2}), we get:
\begin{eqnarray} 
\dot{H} + 2 H^2 \left(1 - \gamma ~\left(\frac{H}{H_I} \right)^{2 \alpha} \right) & = & 0
\end{eqnarray}
whose solution reads,
\begin{eqnarray} 
H & = & \frac{H_I}{\left( \gamma + (1 - \gamma ) \left( \frac{a}{a_I} \right)^{4 \alpha} \right)^{1/{2 \alpha}}}.
\end{eqnarray}
It is easy to see that when $\gamma =0$ (i.e. $\beta =0$), the above equation gives the usual radiation dominated era.
The deceleration parameter is defined as:
\begin{eqnarray}
q & = & - \frac{\ddot{a} a}{\dot{a}^2}.
\end{eqnarray}
The universe is decelerating when $q>0$ and accelerating when $q<0$. Using Friedmann's equations for flat universe, we get:
\begin{eqnarray}
q & = & \frac{1}{2} \left(1 + 3 w \right) = 1 - 2 \beta \rho^{\alpha}.
\end{eqnarray}
It follows that, 
\begin{itemize}
\item{at small scale $ a <<a_I$} \\
\begin{eqnarray}
q & = & -1 +2  \left(\frac{1}{\gamma} -1 \right) \left(\frac{a}{a_I} \right)^{4 \alpha} 
\end{eqnarray}
\item{at large scale $ a >>a_I$} \\
\begin{eqnarray}
q & = & 1 - 2  \left(\frac{\gamma}{\gamma-1} \right) \left(\frac{a_I}{a} \right)^{4 \alpha}
\end{eqnarray}
\end{itemize} 
The acceleration expansion during inflation ends (i.e. $\ddot{a}=0$) when the expansion rate reaches the value,
\begin{eqnarray}
H_{end} & = & \frac{H_I}{(2 \gamma)^{1/ 2 \alpha}}. 
\label{hend}
\end{eqnarray}
To find the evolution of the temperature, we consider the thermodynamics equation,
\begin{eqnarray}
\frac{d p}{d T} & = & \frac{\rho + p}{T} 
\end{eqnarray}
which can be derived from the first principle of thermodynamics \cite{weinberg1972}. 
By using the equations (\ref{eos1}) and (\ref{fr1}), the above thermodynamics equation can be integrated into
\begin{eqnarray}
\frac{T}{T_{I}} & = & \sqrt{\frac{H}{H_I}} ~\left( \frac{1- \gamma \left(\frac{H}{H_I} \right)^{2 \alpha}}{1- \gamma}\right)^{1+ \frac{3}{4 \alpha}} 
\label{temp}
\end{eqnarray}
where $T_I = T (a_I)$ is a constant of integration and corresponds to the temperature at inflation. This result with (\ref{hend}) implies that the temperature at the end of inflation is,
\begin{eqnarray}
T_{end} & = & \frac{T_I}{(2 \gamma)^{1+ \frac{1}{\alpha}} (1- \gamma)^{1+ \frac{3}{4 \alpha}}} 
\end{eqnarray} 

One of the important way to check the causality of the universe to survive is using the speed of the sound. For the causality condition, the speed of the sound must be less than the local light speed, $c_{s}\leq 1$. Moreover, to find the classical stability condition of our model, we use the squared speed of sound $c_s^2$. A positive value of $c_s^2$ represents a stable model whereas a negative value of $c_s^2$ indicates the instability of the model.
 The squared of the sound speed can be calculated as follows \cite{kruglov2015}; 
\begin{equation}
c_{s}^{2}=\frac{dp}{d\rho }=\frac{1}{3} - \frac{4}{3} \beta (1+ \alpha) ~\rho^{\alpha}
\label{cs1}
\end{equation}
which can be expressed nicely as:
\begin{equation}
c_{s}^{2}= \frac{1- \gamma ~2^{\alpha} ~(3 + 4 \alpha) \left( \frac{B^2}{\rho_I}\right)^{\alpha}}{3 ~(1 + \gamma~ 2^{\alpha}~ \left( \frac{B^2}{\rho_I}\right)^{\alpha})} ~~ .
\end{equation}%
The causality and classical stability occurs if the speed of sound varies between in the range $0 \le c_s^2 \le 1$. Figure 3 shows the evolution of sound $c_s^2$ versus $B^2/rho_I$ for different values of the model's parameters $\gamma$ and $\alpha$. One sees that  the causality and classical stability is satisfied for small values of $\gamma$.
\begin{figure}[htp]
\centering
  \includegraphics[width=.6\textwidth]{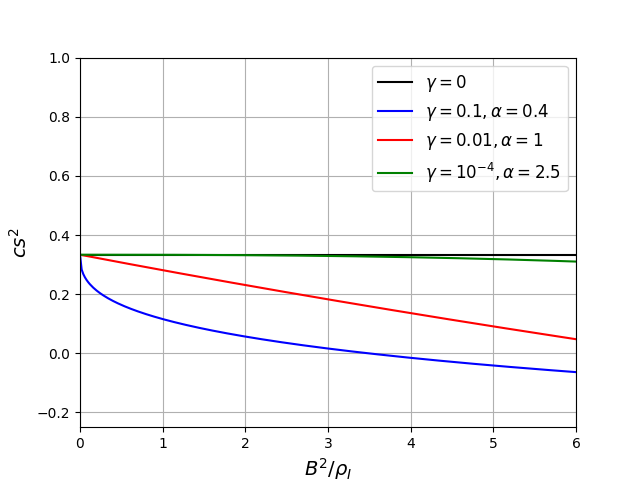} 
\caption{
Variation of the speed of sound $c_s^2$ as a function of $B^2/\rho_I$ for different values of $\gamma$ and $\alpha$.}
\end{figure}

\section{Gravitational Baryogenesis}
Gravitational baryogenesis is one of the mechanism that has been proposed to generate a baryon anti-baryon asymmetry. It is based on a CP-violating interaction between the derivative of the Ricci scalar curvature and the baryonic matter current which has the form \cite{davoudiasl2004}:
\begin{eqnarray}  
\frac{1}{M_{\ast}^2} \int \sqrt{-g} \partial_{\mu} R J_B^{\mu}
\end{eqnarray}
where $M_{\ast}$ is a cut-off mass scale characterizing the effective gravitational theory. Such an interaction term may result from higher order interactions in the fundamental gravitational theory 
\cite{davoudiasl2004}. 
The resulting baryon to entropy ratio for the CP violating interaction is: 
\begin{eqnarray}
\eta_B & \simeq & - \frac{15 g_b}{4 \pi^2 g_{\ast}} \frac{\dot{R}}{{M_{\ast}^2} T} \Big|_{T=T_D}
\label{basym}
\end{eqnarray}

where $T_D$ denotes the decoupling temperature when the baryon current violation decouples, $g_b$ is the number of intrinsic degrees of freedom of the baryons and $g_{\ast}$ is the total number of degrees of freedom. \\
In our model, the Ricci scalar $R$ can be written as: 
\begin{eqnarray}
R & = &   \kappa^2 \left( \rho - 3 p \right) = 4 \kappa^2 \beta \rho^{\alpha+ 1}
\end{eqnarray}
and the derivative of the Ricci scalar with respect to the cosmic time is found to be,
\begin{eqnarray}
\dot{R} & = & 4 \beta (1+ \alpha) \kappa^2 \rho^{\alpha} \dot{\rho} \nonumber \\
& = & - 48 \gamma ~(1+ \alpha)  H_I^3 \left( \frac{H}{H_I} \right)^{2 \alpha +3} \left(1 - \gamma ~\left(\frac{H}{H_I} \right)^{2 \alpha} \right)
\end{eqnarray}
The above equation expresses a close relation between the non-linearity parameter $\beta$ and the net baryon asymmetry which implies that for a radiation dominated Universe $\beta = 0$ (no non-linearity), the derivative of the Ricci scalar is zero, thereby leading to a vanishing baryon asymmetry. \\
To determine the amount of baryon asymmetry, we have to evaluate the decoupling temperature $T_D$ which can be obtained from equation (\ref{temp}), 
\begin{eqnarray}
\frac{T_D}{T_{I}} & = & \sqrt{\frac{H_D}{H_I}} ~\left( \frac{1- \gamma \left(\frac{H_D}{H_I} \right)^{2 \alpha}}{1- \gamma}\right)^{1+ \frac{3}{4 \alpha}} 
\end{eqnarray}
where $H_D = H (a_D)$ is the Hubble parameter at the decoupling scale factor. 
Inserting everything into (\ref{basym}), the resulting baryon asymmetry becomes, 
\begin{widetext}
\begin{eqnarray}
\eta_B & \simeq & \frac{180 g_b}{\pi^2 g_{\ast}} (1+\alpha) ~\gamma (1- \gamma)^{1 + \frac{3}{4 \alpha}} 
\frac{H_I^3}{M_{\ast}^2 T_I} 
\left(\frac{H_D}{H_I} \right)^{2 \alpha + 5/2} ~\left( \frac{1- \gamma \left(\frac{H_D}{H_I} \right)^{2 \alpha}}{1- \gamma}\right)^{- \frac{3}{4 \alpha}}
\end{eqnarray}
\end{widetext}
We proceed by investigating under which conditions the resulting baryon asymmetry is compatible with the observed value. We assume that the cuttoff $M_{\ast}$ is equal to $M_{\ast}=M_p/\sqrt{8 \pi}$ and $T_D = 2 \times 10^{16} ~GeV$ which is compatible with the upper bound of the tensor-mode fluctuations constraints on the inflationary scale $M_I < 3.3 \times 10^{16} ~GeV$ \cite{davoudiasl2004}. Also, we take $g_b \sim O (1)$ and $g_{\ast} \sim 106$ which is the total number of the effectively massless particles in the early Universe.
Results are reported in the figures 4 and 5. Figure 4 shows the variation of the baryon asymmetry as a function of $\gamma$ for a wide range of values of the paramater $\alpha$ whereas figure 5 shows the resulting baryon asymmetry as a function of $\alpha$ for fixed values of $\gamma$. The horizontal dahsed lines denote the phenomenologically allowed region of the observed baryon asymmetry. The main conclusion from these figures is that one can obtain a successful baryogenesis for a wide range values of the model parameter space.
 \begin{figure}[htp]
\centering
  \includegraphics[width=.6\textwidth]{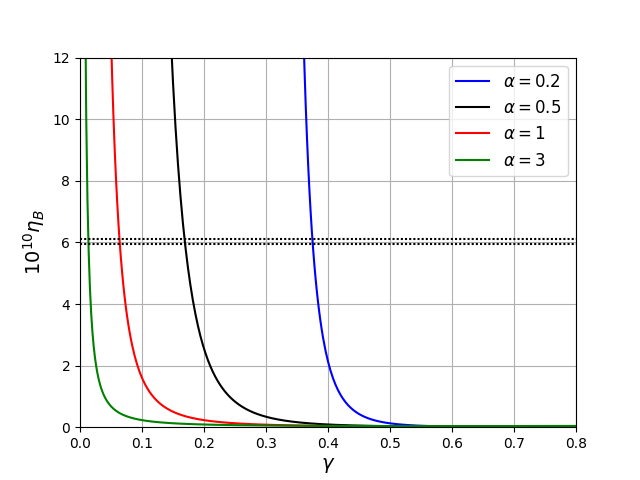} 
\caption{Baryon asymmetry $\eta_B$ versus $\gamma$ for different values of $\alpha$.}
\end{figure}
 \begin{figure}[htp]
\centering
  \includegraphics[width=.6\textwidth]{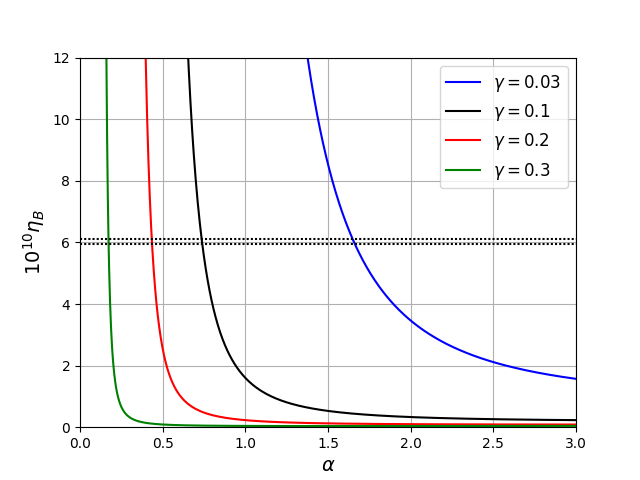} 
\caption{Baryon asymmetry $\eta_B$ versus $\alpha$ for different values of $\gamma$.}
\end{figure}
\section{Conclusion}
In this paper, we presented an economical model that addresses some open cosmological problems, namely, absence of the initial singularity, an early acceleration i.e. inflation and matter-antimatter asymmetry in the Universe. In the early Universe, inflation can be caused by the effect of the very strong non-linear electrodynamics fields. With these in mind, we proposed a non-singular model of the Universe in the framework of general relativity with a non-linear electrodynamics Lagrangian characterized by two parameters $\alpha$ and $\beta$. The usual electrodynamics lagrangian is recovered for $\beta=0$. We showed that such a model leads to an equation of state $p=\frac{1}{3} \rho - \frac{4}{3} \beta \rho^{1+\alpha}$ where the Universe undergoes an early epoch inflation to radiation dominated Universe. The model has the interesting features that there are no singularities of the energy density, pressure, the Ricci tensor, the Ricci tensor squared and the Kretschmann scalar.  Note that one can also consider perfect fluid in the cosmological scenerio to describe the inflation similar to our model \cite{Nojiri:2004pf,Astashenok:2012kb}. However, the important thing is to understand what kind of matter drives the universe acceleration. Here, we specify this matter to be new nonlinear electrodynamics model that leads to inflationary universe.
In order to obtain the baryon asymmetry, we have explicitly calculated the derivative of the Ricci scalar $R$. We have shown that the baryon asymmetry is generated by the non-linear parameter $\beta$. Finally we have calculated the resulting baryon asymmetry and have discussed how a successful gravitational baryogenesis is obtained by comparing our results with the observational data. \\

A full analysis of the inflation in our non-linear electrodynamics model is under progress. Spectral index observables of the model will be calculated and the predicted new spectral index will be compared to some typical inflationary models such as Higgs inflation and its variants where the inflation is driven by the Higgs \cite{Bezrukov2008} and/or extra scalar fields.
\section*{ACKNOWLEDGMENTS}
HBB gratefully acknowledges the financial support from University of Sharjah (grant number V.C.R.G./R.438/2020).

\end{document}